\documentclass[12pt, letterpaper]{article}

\usepackage{amsmath}
\usepackage{amssymb}
\usepackage{amsfonts}
\usepackage{array}
\usepackage{bm}
\usepackage{graphicx}
\usepackage{booktabs}
\usepackage{float}
\usepackage{amsthm}     

\usepackage[authoryear]{natbib}

\title{Genetic Algorithms in Regression}

\author{
	Mo Li\\
	Department of Mathematics,\\
	University of Louisiana at Lafayette,\\
	\texttt{mo.li@louisiana.edu}\\
	\and
	QiQi Lu\\
	Department of Statistical Sciences and Operations Research,\\
	Virginia Commonwealth University,\\
	\texttt{qlu2@vcu.edu}\\
	\and
	Robert Lund\\
	Department of Statistics,\\
	University of California, Santa Cruz,\\
	\texttt{rolund@ucsc.edu}\\
	\and
	Xueheng Shi\\
	Department of Statistics,\\
	University of Nebraska-Lincoln,\\
	\texttt{shixueheng@gmail.com}
}

\begin{document}

\maketitle

\begin{abstract}
Many statistical problems involve optimization over a discrete parameter space having an unknown dimension. In such settings, gradient-based methods often fail due to the non-differentiability of the objective function or a non-convex or massive search space with an objective function having many local maxima/minima. This paper presents \textbf{GAReg}, a unified genetic algorithm package that handles discrete optimization regression problems, which works well when standard algorithms are unjustified. \textbf{GAReg} provides a compact chromosome representation supporting optimal knot placement for regression splines, best-subset regression variable selection, and related problems. The package allows for uniform initialization, constraint-preserving crossover and mutation, steady-state replacement, and an optional island-model parallelization. \textbf{GAReg} efficiently searches high-dimensional model spaces, providing near-optimal solutions in settings where exhaustive enumeration or integer or dynamic programming approaches are infeasible.
\end{abstract}

\section{Introduction}

Many regression problems require optimizing an objective function over a discrete model space of an unknown dimension. This problem arises when determining the optimal number and locations of knots in a spline and selecting the best subset of regression factors. These problems are often framed through minimization of a penalized likelihood having the form
\[
Q(\boldsymbol{\beta}) = -2 \ln(L(\boldsymbol{\beta})) + P(\boldsymbol{\beta}).
\]
Here, $L(\boldsymbol{\beta})$ is the statistical likelihood of a model having the parameter vector $\boldsymbol{\beta}$ and $P(\boldsymbol{\beta})$ is a model complexity penalty such as the Akaike Information Criterion (AIC; \citealp{Akaike1974}) or the Bayesian Information Criterion (BIC; \citealp{Schwarz1978}). Other penalties are possible, including the Minimum Description Length (MDL; \citealp{Rissanen1978}; \citealp{Rissanen2007}), the modified BIC (mBIC; \citealp{Bogdan2004}), and the Extended BIC (EBIC; \citealp{ChenChen2008}), among others. Because candidate models have different dimensions, some optimization problems cannot be addressed with standard gradient-based techniques. Dynamic or integer programming can optimize some nuanced objective functions \citep{killick2012optimal}; however, the structure of the problem often precludes their use. Despite recent advances, mixed-integer programming algorithms \citep{Bertsimas-2016-BSS} remain challenged in ultra-high-dimensional spaces.

Genetic algorithms (GAs) offer an appealing alternative, providing a stochastic search strategy that naturally adapts to discrete, non-differentiable, and combinatorial model spaces. By iteratively improving candidate solutions through selection, crossover, mutation, and generational replacement, GAs capably identify approximate global minimizers of $Q(\boldsymbol{\beta})$, even when the search space is too large to be efficiently handled by other methods. Prior research has demonstrated GA effectiveness for knot selection in regression splines \citep{Lee2002} and multiple changepoint configuration estimation in time series \citep{davis2006structural, lu2010mdl}.

While methodological optimization advancements abound, software support for discrete optimization regression problems remains limited. The \textsf{R} package \textbf{kofnGA} \citep{Scrucca2017} assumes a fixed number of knots and does not allow situations where the number of knots varies.  No general-purpose \textsf{R} package that jointly optimizes both the number and locations of knots in regression splines, or that performs optimal knot placement when the number of knots is fixed, currently exists.

This paper introduces \textbf{GAReg}, an \textsf{R} package that implements a GA for discrete optimization in statistical regression settings. \textbf{GAReg} supports multiple tasks, including best regression subset selection and optimal knot number selection and placement in splines.  Problem-specific constraints such as minimum spacings restrictions are accommodated. 

In spline problems, \textbf{GAReg} optimizes both the number and locations of knots using a penalized Gaussian likelihood, providing substantial practical value. A minimum distance constraint between consecutive knots is allowed. For variable selection problems, our best subset regression GA uses a binary chromosome representation adapted from \textsf{R} package \textbf{GA}. Here, \textbf{GAReg} leverages the \textsf{R} packages \textbf{GA} and \textbf{changepointGA} engines to supply task-specific objective functions.

\textbf{GAReg} employs uniform chromosome initialization, constraint-preserving breeding and mutation, and a steady-state replacement scheme. An optional island-model parallelization improves search space exploration and reduces premature convergence. The package efficiently explores high-dimensional discrete spaces and offers near-optimal solutions in settings where traditional optimization methods are infeasible or perform poorly.

The remainder of the paper is organized as follows. Section 2 introduces the GA framework and our chromosome representations. Section 3 describes the package's architecture and the user-level interface for spline fitting and best regression subset selection. Section 4 presents illustrative examples detailing optimal knot placement in regression splines, joinpoint regression fits, and best subset variable selection. Section 5 concludes with a summary of the main features of the package.

\section{Genetic Algorithms}

A genetic algorithm (GA) is a stochastic search procedure that iteratively improves a collection (population) of candidate models (solutions) by imitating several ``survival of the fittest'' mechanisms inherent in natural selection processes. GAs represent each candidate model as a chromosome, beginning with a population of $n$ models, viewed collectively as the initial generation. GAs iteratively produce generations of new models (usually also of size $n$), with each generation having improved fitness characteristics (better minimize $Q$). When done strategically, the best fitting member(s) of the current generation, which is the model(s) in the current generation that minimizes $Q$, converges to the optimal model. In short, GAs are intelligent stochastic searches that bypass the need to exhaustively evaluate all possible models.

Our GAs use the following genetic operators: 1) \emph{selection}, which stochastically favors individuals with lower $Q$ scores to breed children in the next generation. Specifically, the current generation members are ranked from $0$ to $n_{\rm gen}-1$ in their fitness level of minimizing $Q$ ($0$ being the least fit, $n_{\rm gen}-1$ the most fit; ranks are split for any ties) and parents of the next generation's children are chosen in inverse proportion to these ranks. 2) \emph{crossover} combines two parental chromosomes (mother and father) in some manner to breed/produce a child chromosome. 3) low-probability \emph{mutation} perturbs child chromosomes, ensuring continual exploration of different and possibly new model regions, preventing convergence to a local (but possibly not global) minima of the objective function. 

Our GA adopts a steady–state replacement scheme (Davis, 1991): a newly created child replaces the population’s least–fit member only if it improves fitness and is not a duplicate of another current generation member; otherwise, it is discarded and another child is independently generated. Since replacements require improvement, the minimal $Q$-score decreases over successive generations. The algorithm terminates when some stopping criterion is met (e.g., convergence is achieved or a computational budget is exhausted), and the chromosome with the lowest $Q$ score in the final generation is returned as the approximate minimizer of $Q(\boldsymbol{\beta})$. The associated estimate $\hat{\boldsymbol{\beta}}$ is the model represented by this chromosome. 

A widely used GA variant is an island scheme, which partitions the total current population into several isolated subpopulations (termed islands) that largely evolve independently. Occasional island to island migration, which enhances exploration and reduces premature convergence (see \cite{lu2010mdl, li2024changepointga} for details), is allowed.


We first describe optimal spline knot selection to the independent data pairs $(x_i, Y_i)_{i=1}^n$. Here, candidate models are encoded via a chromosome $C$ of form
\begin{equation}
	C = (m; \tau_1, \ldots, \tau_m)^\top,
	\label{eq:chromosome}
\end{equation}
where $m$ is a candidate dimension and $\boldsymbol{\tau}=(\tau_1, \ldots, \tau_m)^\top$ indexes the $m$ knot locations. For example, in a series of length $n=100$, $C=(3; 11, 43, 86)^\top$ represents a model having three knots at the locations $x_{11}, x_{43}$, and $x_{86}$. For ease of further exposition, we assume that $\{ x_i \}_{i=1}^n$ are distinct and ordered ($x_1 < x_2 < \ldots < x_n$) (\textbf{GAReg} can handle replicate or unordered design points). Our package requires all interior knot locations to be one of the interior $x_i$s, which aligns with standard B-spline practice. The user also sets a minimum spacing between consecutive knots, denoted by $d_{\rm min}$, which is enforced in the fitting.

Considering the fixed-$m$ case for the moment, the knot location vector $\boldsymbol{\tau}$ contains $m$ distinct interior knot indices satisfying the minimum spacing requirements. Given these ordered $m$ knot locations, say $\boldsymbol{\tau}=(x_{i_1}, \ldots, x_{i_m})^\top$, \textbf{GAReg} fits the spline regression model 
\begin{equation}
	\label{JP}
	\boldsymbol{Y} = \boldsymbol{X}(\boldsymbol{\tau}) \boldsymbol{\beta}
	(\boldsymbol{\tau})
	+ \boldsymbol{\epsilon},
\end{equation}
where $\bm{Y}=(Y_1, \ldots, Y_n )^\top$ is the response vector, $\bm{X}(\bm{\tau})$ is the $n \times p$ regression spline design matrix, $\bm{\beta}(\bm{\tau})$ is the parameter vector (whose dimension and entries may depend on $\bm{\tau}$), and $\bm{\epsilon}=(\epsilon_1, \ldots, \epsilon_n)^\top$ contains the model errors. \textbf{GAReg} determines the design matrix $X(\bm{\tau})$, determines the best estimate of $\bm{\beta}$, and scores the fit with a Gaussian likelihood and the BIC penalty $P(\bm{\beta}) = r \ln(n)$; here $r$ is the number of free model parameters in the spline fit. Additional details are given in Section 4.

To produce the initial generation of chromosomes, we need to sample $n_{\rm gen}$ knot location vectors $\bm{\tau}$, each having length $m < n_{\rm gen}$. To sample a single chromosome, we generate a random sample of $m$ non-decreasing and non-replicated integers in $\{ 1, \ldots, n \}$; call these $i_1 < i_2 < \cdots < i_m$. Should $x_{i_j} - x_{i_{j-1}} > d_{\rm min}$ for each $j \in \{ 2, \ldots, m \}$, the minimal spacing criterion is satisfied and the $m$ knots are placed at the design points $x_{i_1}, \ldots, x_{i_m}$ and $\bm{\tau}=(i_1, \ldots, i_m)^\top$. The boundary conditions $x_{i_1} > x_1 + d_{\rm min}$ and $x_{i_m} < x_n - d_{\rm min}$ are imposed. Should this generation procedure encounter an impossibly at any stage --- that is, the minimal spacing requirements cannot be met --- then the chromosome is discarded altogether and the generation procedure is started from scratch. As $n_{\rm gen}$ chromosomes are produced, checks for duplicates are made, ensuring that no two chromosomes are identical. In the end, the procedure produces a random sample of $n_{\rm gen}$ equally likely chromosome knot locations that satisfy the spacing requirements and do not duplicate one and other.

For chromosome breeding in subsequent generations, two parents are first selected by a linear ranking mechanism. Specifically, the $n_{\rm gen}$ chromosomes are ranked from $0$ to $n_{\rm gen}-1$ in order of their $Q$ scores (rank $0$ is assigned to the largest $Q$ score, rank $n_{\rm gen}-1$ to the smallest $Q$ score; ties split ranks). The first parent (say the father) is chosen as the $k$th chromosome with probability
\[
\frac{2k}{(n_{\rm gen}-1) n_{\rm gen}}.
\]
The second parent (say the mother) is then chosen in the same ``ranked'' manner from the remaining $n_{gen}-1$ chromosomes. This procedure produces children that stochastically favor lower $Q$ values (the fittest), while retaining population diversity \citep{davis2006structural, lu2010mdl}. 

To describe our crossover chromosome breeding mechanism, denote the two length-$m$ parent chromosomes by $\boldsymbol{\tau}^{\text{Mom}}=(\tau_{1}^{\text{Mom}}, \ldots, \tau_{m}^{\text{Mom}})$ and $\boldsymbol{\tau}^{\text{Dad}}=(\tau_{1}^{\text{Dad}}, \ldots, \tau_{m}^{\text{Dad}})$.   The chromosome of the child to be produced is denoted by $\boldsymbol{\tau}^{\text{Child}}=(\tau_{1}^{\text{Child}}, \ldots, \tau_{m}^{\text{Child}})$ and is generated as follows. Our algorithm proceeds sequentially in $i$. At $i=1$, $\tau_{1}^{\text{Child}}$ is drawn with probability 1/2 from the two values $\tau^{\text{Dad}}_1$ and $\tau^{\text{Mom}}_1$. For each index $i \in \{ 2, \ldots, m \}$, we only consider parental knots satisfying the minimal spacing condition.  This is done by defining a set of ``next knot location indices" $C_i$ that can contain at most two elements.  In particular, $C_i$ contains both $\tau_i^{\text{Dad}}$ and $\tau_i^{\text{Mom}}$ if both parents satisfy the minimal spacing requirements: $x_{{\tau_i}^{\rm Dad}}-x_{{\tau_{i-1}^{\rm Child}}} > d_{\rm min}$ and $x_{{\tau_i}^{\rm Mom}}-x_{{\tau_{i-1}^{\rm Child}}} > d_{\rm min}$. In this case, one of these two is chosen at random.  In the case where $C_i$ has unit cardinality and one of the spacing restrictions is violated, that unique candidate is taken. Should $C_i$ be empty, the partially constructed child chromosome is discarded and the procedure restarts from $i=1$. After this child is generated, $\boldsymbol{\tau}_{\rm Child}$ is additionally checked for parent cloning: if $\boldsymbol{\tau}^{\text{Child}}=\boldsymbol{\tau}^{\text{Dad}}$ or $\boldsymbol{\tau}^{\text{Child}}=\boldsymbol{\tau}^{\text{Mom}}$, the child is rejected and the construction restarts from scratch. To guarantee termination, we cap the number of restart attempts; if this cap is exceeded, $\boldsymbol{\tau}^{\text{Child}}$ is generated via our above uniform sampler.

Our mutation mechanism entails a global refresh rather than any chromosome perturbation. Specifically, when a mutation occurs, which is set to have probability 0.3 in our program, the child's chromosome is replaced by a new draw from the same sampler used to produce the initial generation. The mutation probability can be set by the user. Our mutation injects diversity into the population by allowing occasional ``long-range moves" to new unexplored model space regions, helping the search escape any local minima and reduce the premature convergence. 

Replacement is steady–state in both modes: a child enters the population only if it strictly improves fitness and is not a duplicate of any other member.  As such, the best $Q$-score is non-increasing across accepted replacements (increasing generations).

When $m$ is allowed to vary, our GA settings default to those in \textbf{changepointGA}; see \cite{lu2010mdl} and \cite{li2024changepointga} for details. The chromosomes are still required to satisfy our constraints. Parent selection, crossover, mutation, and steady-state replacement are as described in \textbf{changepointGA}; \textbf{GAReg} does not alter these operators.

For all knot selection tasks, an optional island model GA is supported. This option allows equal-sized subpopulations that evolve largely independently; sporadic swapping of inhabitants is periodically done at random. Island GAs tend to enhance model exploration and reduce premature convergence.

Moving to regression variable selection problems, the optimal number of variables is bounded by the number of possible regressors, which we denote by $p$. In this case, the order of the regressors is irrelevant, and there are usually no regression factor minimal spacing restrictions. Unlike optimal knot placement, which is a sparse location search on an ordered axis, variable selection is a combinatorial subset search over unordered predictors. Accordingly, we delegate the genetic search to the more general-purpose \textsf{R} package \textbf{GA} engine. 

In this architecture, binary chromosome representations of length $p$ are used: 
\[
C=(z_1, \ldots, z_p)^\top,
\]
where $z_j \in \{ 0, 1 \}$ for each $j \in \{ 1 , \ldots , p \}$. Here, $z_j=1$ indicates inclusion of regressand $j$ and $z_j=0$ its exclusion.

Chromosome initialization, selection, crossover, mutation, and replacement follow \textbf{GA}'s default operators and computation path for binary strings. Our \textbf{GAReg} supplies the objective function (Gaussian) with the BIC penalty $P(\boldsymbol{\beta})= r \ln(n)$, where $r=\sum_{j=1} z_j$ is the number of regressands in the model. This penalty is understood to be zero should $r=0$.

\section{Package Architecture}

\textbf{GAReg} has two drivers covering the package’s core tasks. We first describe \verb|gareg_knots()| for optimal spline-knot placement under either ``fixed-$m$" or ``varying-$m$" modes. Thereafter, \verb|gareg_subset()| for best‐subset regression is described. Implementation details are illustrated with worked examples in the next section.

\subsection{\textit{gareg\_knots()} (fixed-$m$ vs.~varying optimal knot placements)}

Given responses $\{ Y_i \}_{i=1}^n$ and design values $\{ x_i \}_{i=1}^n$, the interior knot locations, and optionally their number, to optimize a user-specified criterion for a spline regression of $Y_i$ on the $x_i$s. The following function call supports (i) a fixed-$m$ knot count that has exactly $m$ distinct knots, and (ii) a varying-$m$ knot count that searches jointly over the number of interior knots and their locations, minimizing a BIC penalized Gaussian likelihood.
\begin{verbatim}
	gareg_knots(y, x, ObjFunc=NULL, fixedknots=NULL, minDist=3L, degree=3L, 
	type = c("ppolys", "ns", "bs"), intercept=TRUE, gaMethod="cptga", 
	cptgactrl=NULL, monitoring=FALSE, seed=NULL, ...)
\end{verbatim}

\noindent\textit{Key arguments:}
\begin{itemize}
	
	\item \textbf{Core data:} \texttt{y} (response), \texttt{x} (index or covariate; defaults to \texttt{seq\_along(y)}). A candidate grid $\{x_{(1)} < \ldots < x_{(m)}\}$ is formed from the sorted unique values of the $x_i$s unless supplied via \texttt{x\_unique} inside the objective functions.
	
	\item \textbf{Mode selection:} \texttt{fixedknots}. Set \texttt{fixedknots=NULL} (default) to enable varying-$m$ mode and estimate both the number of interior knots $m$ and their locations. Given a user-specified $m$, \texttt{fixedknots=m} searches only over configurations having exactly $m$ interior knots.
	
	\item \textbf{Spline specification:} choose \verb|type| $\in \{\verb|"ppolys"|, \verb|"ns"|, \verb|"bs"|\}$, and set the integer \verb|degree| (required for \verb|"ppolys"| and \verb|"bs"|; ignored for \verb|"ns"|) and the logical \verb|intercept| (whether to include a constant column). 
	
	\begin{itemize}
		\item \verb|"ppolys"| (piecewise polynomials): degree-$d$ truncated–power regression splines; \verb|degree| (required) sets $d$, and \verb|intercept| is optional.
		\item \verb|"ns"| (natural cubic): cubic spline with natural boundary conditions; \verb|degree| is ignored, and \verb|intercept| is optional.
		\item \verb|"bs"| (B-spline): degree-$d$ B-spline basis with compact local support and good numerical conditioning; \verb|degree| (required) sets $d$, and \verb|intercept| is optional.
	\end{itemize}
	
	\item \textbf{Search constraints:} \texttt{minDist} enforces a minimum spacing between adjacent interior knot locations, preventing tightly clustered knots.
	
	\item \textbf{Objective function:} \verb|ObjFunc| defines the fitness to be \emph{minimized}. If omitted, \verb|gareg_knots| defaults to an information criterion objective: \verb|varyknotsIC| in varying-$m$ mode and \verb|fixknotsIC| in fixed-$m$ mode. These evaluate BIC/AIC/AICc from a spline fit. Users may supply a custom objective of the form \verb|f(chromosome, y, x, x_unique, ...)|; additional covariates (e.g., \verb|x_base|) and other possible tuning parameters can be passed via \verb|...|. The objective returns a single scalar (smaller is better), with non-feasible inputs mapped to \verb|Inf|.
	
	\item \textbf{Engine choice and controls:} delegates to \textbf{changepointGA} package. \texttt{gaMethod} selects the backend (\texttt{``cptga''} for single population GA model or \texttt{``cptgaisl''} for an island GA model). The engine controls (population size, crossover, mutation rates, island parameters, iteration limits, etc.) and are set through \texttt{cptgaControl()} or a named list; \texttt{monitoring} toggles progress output; \texttt{seed} fixes RNG state. In the fixed–$m$ mode, feasibility-preserving operators are injected unless explicitly overridden.
\end{itemize}

\noindent\textit{Use guidance:} 

Choose \texttt{fixedknots} when domain knowledge suggests a specific $m$; otherwise, use the varying–$m$ default and rely on an information criterion to balance fit and knt numbers. Set \texttt{minDist} on the \texttt{x} grid to have $d_{\rm min}={\rm minDist}$. For custom objectives, ensure they return finite values on feasible chromosomes and $+\infty$ (or some suitably large penalty) otherwise. The return value is an S4 \texttt{``gareg''} object that records the best fitness, the selected knot indices, and the full fit; \texttt{summary()} reports the optimal configuration.

\subsection{\textit{gareg\_subset()} (variable selection)}
Given responses $\{ Y_i \}_{i=1}^n$ and a design matrix $\bf{X}$, we want to perform a GA-based best-subset variable selection by optimizing a user objective with a binary chromosome. The following function call supports the task; some arguments are omitted to avoid redundancies.

\begin{verbatim}
	gareg_subset(y, X, ObjFunc = NULL, gaMethod = "ga", gacontrol = NULL, monitoring = FALSE,
	seed = NULL, ...)
\end{verbatim}

\noindent\textit{Key arguments:}

\begin{itemize}
	
	\item \textbf{Data inputs:} \verb|y| (length $n$) and \verb|X| ($n\times p$ matrix of candidate predictors).
	
	\item \textbf{Chromosome encoding:} Binary chromosome over the $p$ predictors with 1 indicates inclusion.
	
	\item \textbf{Engine selection:} \verb|gaMethod| chooses package \textbf{GA} engine via \verb|ga()| for the single population GA model or \verb|gaisl()| for the island GA model.
	
	\item \textbf{Engine controls:} \verb|gacontrol| passes GA hyperparameters to package \textbf{GA} (e.g., \verb|popSize|, \verb|maxiter|, \verb|run|, \verb|pcrossover|, \verb|pmutation|, \verb|elitism|, \verb|seed|, parallel/islands settings). Model-specific options (e.g., \verb|family|, \verb|weights|, \verb|offset|, \verb|control|) are forwarded via \verb|...| to the objective, avoiding name clashes with engine fields.
	
	\item \textbf{Default fitness:} \verb|subsetBIC| supplies the fitness as the \emph{negative} BIC, so larger values are better and the GA runs with \verb|max = TRUE|. Concretely, BIC is computed as $n \ln(\mathrm{RSS}/n) + k\ln(n)$ for Gaussian models with the identity link (using the residual sum of squares, RSS), and with the residual deviance used in place of RSS for generalized linear models; \verb|subsetBIC| returns its negative, \(-\mathrm{BIC}\). Users may instead pass any objective function that accepts the binary chromosome as its first argument; if the proposed objective is a loss to be minimized, either return its negative or set \verb|gacontrol=list(max = FALSE)| to switch the engine to minimization.
\end{itemize}

\noindent\textit{Use guidance:} 
Start with \texttt{subsetBIC} for Gaussian/GLM models; switch to a custom scoring rule with a domain-specific loss or with constraints. The output is an S4 \verb|"gareg"| object that contains the GA fit, best fitness, and subset, summarized by the number of selected variables and their indices. The chosen subset can be refit to obtain $\hat{\boldsymbol{\beta}}$. Users can also use \texttt{summary()} to better report the chosen subset.

\section{Examples and Further Guidance}

\subsection{Optimal Spline Knot Placement} 
\label{subsec::opknotsspline}
Let $(x_i, Y_i)_{i=1}^n$ be bivariate data points, with $x_i$ being the $i$th deterministic abscissa and $Y_i$ the $i$th random ordinate. A $d$th degree regression spline is a piecewise polynomial of degree $d$ that is continuous and has continuous derivatives up to order $d-1$ at each knot. For a sequence of $m$ knots at the ordered locations $x_{\tau_1} < x_{\tau_2} < \cdots < x_{\tau_m} < n$, a degree-$d$ regression spline fits the basis functions $\{ b_j(x) \}_{j=1}^{m+d}$ in the representation
\[
Y_i = \beta_0 + \beta_1 b_1(x) +  \cdots + \beta_d b_d(x)+\sum_{k=1}^m\beta_{d+k} b_{d+k}(x) + \epsilon_i, 
\]
where $b_j(x) = x^j$ for $j \in \{ 1, \ldots, d \}$ and $b_{d+j}(x) = (x - x_{\tau_j})_+^d$ for $j = 1, \ldots, m$, with $x_+ = \max \{ x , 0 \}$.  Here, $\{ \epsilon_i \}_{i=1}^n$ is IID zero mean noise with the variance $\sigma^2$.  When $d = 3$, the setup is called a cubic spline. Should the second derivative be constrained as zero at the boundaries ($x_1$ and $x_n$) of the domain, the resulting spline is called a \emph{natural cubic spline}. The knot locations are all required to be one of the design points $x_i$.

Another widely used alternative is the B-spline, which provides a numerically stable and efficient basis for splines. A B-spline of degree $d$ is a piecewise polynomial of degree $d$ defined recursively by the Cox -- de Boor formula. The zeroth-degree B-splines are indicator functions over the knot intervals:
\[
B_{i,0}(x) =
\begin{cases}
	1, & \text{if}~x_{\tau_i} \le x < x_{\tau_{i+1}}, \\[4pt]
	0, & \text{otherwise}
\end{cases}.
\]
Higher degree B-splines are constructed recursively in $d$ via
\[
B_{i,d}(x) =
\frac{x - x_{\tau_i}}{x_{\tau_{i+d}} - x_{\tau_i}} B_{i,d-1}(x)
+
\frac{x_{\tau_{i+d+1}} - x}{x_{\tau_{i+d+1}} - x_{\tau_{i+1}}} B_{i+1,d-1}(x),
\]
where any divisions by zero are interpreted as zero. Because B-splines have a compact local support, each $B_{i,d}(x)$ is nonzero over at most $d+1$ consecutive knot intervals.  Accordingly, sparse design matrices and excellent numerical stability results. These properties make B-splines particularly appealing for regression modeling, smoothing, and functional data analysis \citep{deBoor2001, Hastie2009}.

The placement of spline knots critically influences the flexibility and smoothness of the spline. Common knot selection strategies include placing knots at predictor quantiles to capture regions of high data density, or to place them at equally-spaced locations across the design range \citep{deBoor2001, Hastie2009}. These heuristic methods do not guarantee optimal knot numbers or knot locations. Careful data-driven knot selection, particularly in the context of smoothing splines, can substantially enhance model fit and interpretability \citep{Ruppert2003}.

Given the $m$ knot location indices $\boldsymbol{\tau} = (\tau_1, \tau_2,  \cdots, \tau_m)^\top$, let ${\bf X}(\boldsymbol{\tau})$ denote the associated spline regression design matrix based on the truncated power or B-spline basis function representations. The spline regression has the form in (\ref{JP}), where $\boldsymbol{\epsilon} \sim \boldsymbol{N}({\bf 0}, \sigma^2 {\bf I}_n)$. The ordinary least squares estimate of $\boldsymbol{\beta}(\bm{\tau})$ is $\hat{\boldsymbol{\beta}}(\boldsymbol{\tau})= [{\bf X}(\boldsymbol{\tau})^\top {\bf X}(\boldsymbol{\tau})]^{-1} {\bf X}(\boldsymbol{\tau})^\top{\bf Y}$, which yields the residual sum of squares (RSS)
\[
\mathrm{RSS}(\boldsymbol{\tau}) 
= 
\bigl\| {\bf Y} - X(\boldsymbol{\tau}) \hat{\boldsymbol{\beta}}(\boldsymbol{\tau}) \bigr\|_2^2.
\]
The Bayesian information criterion (BIC) for the knot configuration in $\boldsymbol{\tau}$ is
\[
\mathrm{BIC}(\boldsymbol{\tau})
=
n \ln \left(\tfrac{\mathrm{RSS}(\boldsymbol{\tau})}{n}\right)
+
r \ln(n),
\]
where $r$ denotes the number of free parameters to be estimated. For instance, a cubic regression spline with a truncated-power basis has $r=m+4$ free parameters (including the intercept), while a natural cubic spline with $m$ interior knots has $r=m+2$ parameters.  When both the number and locations of knots are unknown, the optimal configuration can be obtained by minimizing the BIC criterion:
\[
\hat{\boldsymbol{\tau}}
=
{\arg \min}_{\boldsymbol{\tau}}
\left\{
n \ln\left(\tfrac{\mathrm{RSS}(\boldsymbol{\tau})}{n}\right)
+ 
r \ln(n)
\right\}.
\]

While using GAs to fit regression splines was proposed in \cite{Lee2002}, the \textbf{GAReg} package implements a general and efficient framework for optimal knot placement in cubic, natural cubic, and B-spline regression splines. The algorithm supports both fixed-knot configurations, where the number of knots $m$ is specified by the user, or adaptive settings where $m$ is selected via BIC to balance model fit and complexity. 

\begin{figure}[H]
	\centering
	\includegraphics[scale=0.55]{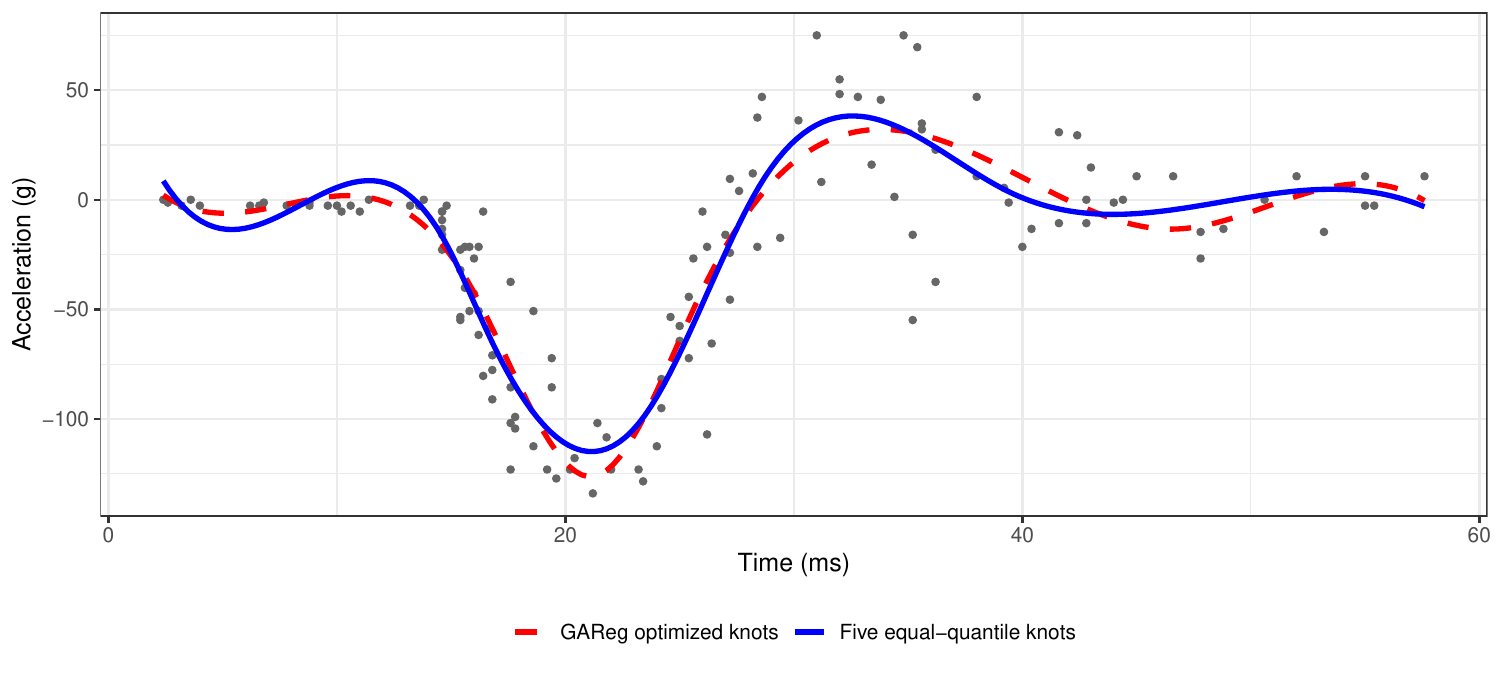}
	\caption{The red dashed curve is a B-spline fit obtained using \texttt{GAReg} optimized knots with the single population GA model (four knots were chosen); the blue solid curve is a spline fit based on five equal-quantile knots.}
	\label{fig:Equal_quant_plot_varying_num_knot}
\end{figure}

To illustrate its application, we consider the \texttt{mcycle} dataset from the \texttt{MASS} package, which records head acceleration measurements from a simulated motorcycle crash used for testing protective helmets. The spline fits against the data are plotted in Figure \ref{fig:Equal_quant_plot_varying_num_knot}.

\begin{verbatim}
	library(MASS)
	data(mcycle)
	y <- mcycle$accel
	x <- mcycle$times
	x_unique <- unique(x)
	
	## GAReg varying knots
	g1 <- gareg_knots(
	y = mcycle$accel, x = mcycle$times,
	minDist = 5,
	gaMethod = "cptga",
	cptgactrl = cptgaControl(popSize = 200, pcrossover = 0.9, pmutation = 0.3, maxgen = 10000),
	ic_method = "BIC"
	)
	bsfit.vary.ga <- lm(y ~ bs(x, knots = x_unique[g1@bestsol], Boundary.knots = range(x)))
	
	## Five equal quantile knots
	place_knots_equal_quantile <- function(x, K) {
		stopifnot(K >= 1)
		x <- sort(na.omit(x))
		# quantile levels: equally spaced within (0,1)
		probs <- seq(0, 1, length.out = K + 2)[-c(1, K + 2)]
		# compute quantiles
		knots <- quantile(x, probs = probs, type = 7)
		as.numeric(knots)
	}
	eq_quant_knots <- place_knots_equal_quantile(x, 5)
	fit_bs_eqquant <- lm(y ~ bs(x, knots = eq_quant_knots, degree = 3))
\end{verbatim}

\subsection{Jointpoint regressions}

When analyzing data with trends such as cancer mortalities \citep{kim-2000-permutation} or surface temperatures \citep{beaulieu-2024-nature}, interest often lies in detecting recent trend changes. Joinpoint regression models, composed of multiple continuous linear segments, are often employed to quantify the scenario. The \textbf{GAReg} package fits joinpoint regressions by providing optimally placed knot locations: when the highest polynomial degree is restricted to one, knot placement reduces to breakpoint detection in a continuous piecewise linear function, commonly referred to as the joinpoint model. 

\begin{figure}[H]
	\centering
	\includegraphics[scale=0.55]{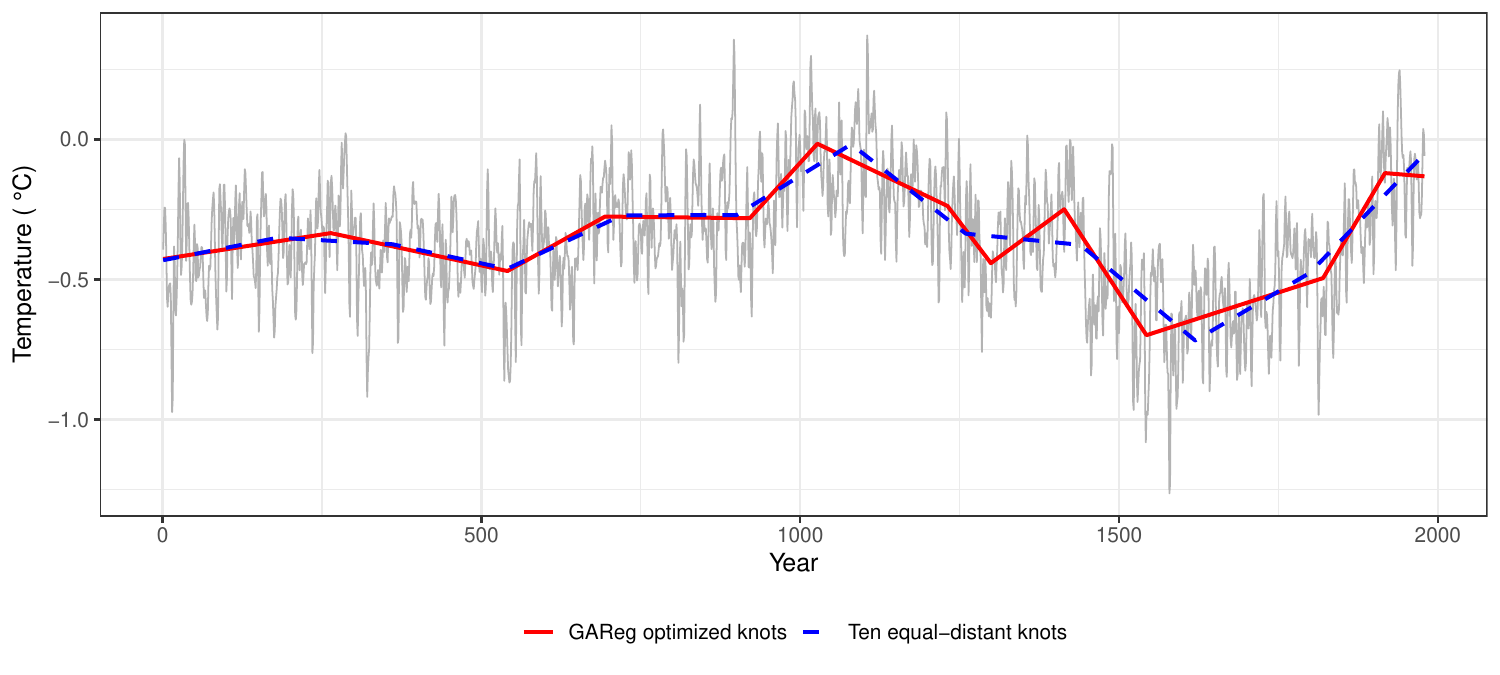}
	\caption{A joinpoint analysis of the Earth's temperatures over the last two thousand years. The red curve depicts a \texttt{GAReg} optimized fit with 11 knots, while the blue curve fit is obtained using 10 equally spaced knots.}
	\label{fig:joinpoints}
\end{figure}

Our next application fits a multiple linear joinpoint model to paleoclimatic temperatures of the Earth over the last two thousand years. See \cite{Moberg-2005-Paleoclimate} for a detailed description of the data. Figure \ref{fig:joinpoints} shows two linear joinpoint fits, one with eleven knots optimized by \texttt{GAReg} and one with ten equally spaced knots. Although both fits are similar and suggest that the Earth has undergone multiple temperature rate changes over the period of record, the \texttt{GAReg}-optimized model fits better, improving our estimate of underlying trend changes.

\begin{verbatim}
	ga.knots <- gareg_knots(
	y = dat$Temp, x = dat$Year,
	minDist = 50,
	gaMethod = "cptgaisl",
	type = "bs", # natural cubic (degree ignored)
	degree = 1, # ignored for 'ns'
	intercept = TRUE,
	cptgactrl = cptgaControl(popSize = 200, pcrossover = 0.9, pmutation = 0.3, 
	maxMig = 100, maxgen = 2000),
	ic_method = "BIC"
	)
	summary(ga.knots) 
\end{verbatim}

\subsection{Best-subset Variable Selection}

Consider independent responses $Y_i$ for $i=1, \ldots, n$, and let $\boldsymbol{x}_i=(x_{i,1}, \ldots, x_{i, p})^\top$ denote the corresponding $p$-dimensional predictor vector. We model $Y_i$ using a generalized linear model (GLM) with link function $g(\cdot)$, so that
\[
\mathbb{E}[Y_i]=g^{-1}\!\left(\boldsymbol{x}_i^\top\boldsymbol{\beta}\right),
\]
where $\boldsymbol{\beta} \in \mathbb{R}^p$ is an unknown coefficient vector. In high-dimensional statistics, it is common to assume that $\boldsymbol{\beta}$ is sparse.

Let $S = \{j_1, \ldots, j_m \} \subseteq \{ 1, \ldots, p \}$ index a candidate subset of $m \leq p$ regressors. Denote by $\hat{\boldsymbol{\beta}}_S$ the estimated coefficient vector under the model restricted to predictors in $S$. The corresponding log-likelihood is
\[
L(S)=\sum_{i=1}^n \log f\!\left(Y_i \mid \boldsymbol{x}_{i,S},\,\hat{\boldsymbol{\beta}}_S\right),
\]
where $f$ is the probability density (or mass) function of $Y_i$ under the assumed GLM, and $\boldsymbol{x}_{i,S}$ denotes the subvector of $\boldsymbol{x}_i$s containing only predictors indexed in $S$. Thus, the BIC model selection amounts to choosing the subset $S$ that minimizes
\[
Q(S)= -2L(S) + m\log(n),
\qquad S \subseteq \{ 1, \ldots, p \}.
\]

Minimizing $Q(S)$ yields a subset $S$ that balances goodness-of-fit with model complexity through a BIC-penalized likelihood criterion. When the number of candidate predictors is small (often $p \le 64$), an exhaustive search over all $2^p$ subsets (best subset regression) can identify the global minimizer. However, the search quickly becomes computationally prohibitive as $p$ increases, since the number of candidate models grows exponentially. Common alternatives include forward and backward stepwise procedures \citep{blanchet2008forward, wang2009forward}, but these algorithms are not guaranteed to recover the optimal subset. This limitation has motivated the use of randomized optimization methods \citep{Wallet-1997-GAselection, Cho-2002-GAselection, Tan-2008-GAselection} as well as mixed-integer programming formulations \citep{Bertsimas-2016-BSS}.

The \textbf{GAReg} framework uses a GA to efficiently explore the high-dimensional model space, providing near-optimal solutions to the best subset regression problem, even when an exhaustive search is computationally infeasible. In this framework, each candidate model is encoded as a chromosome whose binary representation indicates the inclusion or exclusion of predictors. The algorithm iteratively evolves these chromosomes through selection, crossover, and mutation operations, guided by a fitness function based on a BIC penalty. This evolutionary search strategy balances exploration and exploitation, enabling efficient identification of parsimonious models with strong explanatory power.

\begin{verbatim}
	sim_subset_data <- function(n = 60, p = 50, s0 = 25, sigma = 1.5,
	magnitudes_range = c(0.5, 2),
	rho = NULL,
	seed = NULL) {
		stopifnot(n > 0, p > 0, s0 >= 0, s0 <= p, sigma >= 0)
		if (!is.null(seed)) set.seed(seed)
		
		X <- matrix(rnorm(n * p), n, p)
		
		# Active set and coefficients
		true_idx <- if (s0 > 0) sort(sample.int(p, s0)) else integer(0)
		signs <- if (s0 > 0) sample(c(-1, 1), s0, replace = TRUE) else numeric(0)
		magnitudes <- if (s0 > 0) {
			runif(s0, magnitudes_range[1], magnitudes_range[2])
		} else {
			numeric(0)
		}
		
		beta_true <- numeric(p)
		if (s0 > 0) beta_true[true_idx] <- magnitudes * signs
		
		if (is.null(rho)) {
			e <- rnorm(n, sd = sigma)
		} else {
			sd_innov <- sigma * sqrt(1 - rho^2)
			burn_in <- 100
			z <- rnorm(n + burn_in, sd = sd_innov)
			e_full <- numeric(n + burn_in)
			for (t in 2:(n + burn_in)) e_full[t] <- rho * e_full[t - 1] + z[t]
			e <- e_full[(burn_in + 1):(burn_in + n)]
		}
		
		y <- as.numeric(X %*% beta_true + e)
		
		DF <- data.frame(y = y, as.data.frame(X))
		colnames(DF)[-1] <- paste0("X", seq_len(p))
		
		list(
		X = X,
		y = y,
		beta_true = beta_true,
		true_idx = true_idx,
		DF = DF,
		rho = if (is.null(rho)) NULL else rho,
		args = list(
		n = n, p = p, s0 = s0, sigma = sigma,
		magnitudes_range = magnitudes_range,
		rho = rho, seed = seed
		)
		)
	}
	
	sim <- sim_subset_data(n = 100, p = 50, s0 = 25, sigma = 1.5, rho = NULL, seed = 123)
	y <- sim$y
	X <- sim$X
	ga.bbs <- gareg_subset(
	y = y, X = X, gaMethod = "GA", monitor = FALSE,
	gacontrol = list(
	popSize = 120,
	maxiter = 20000,
	run = 4000,
	pmutation = 0.02
	)
	)
	summary(ga.bbs) 
\end{verbatim}

\section{Summary}
This paper introduced an \texttt{R} package that uses genetic algorithms to handle several non-standard regression issues where standard algorithms often fail. The package handles optimal spline knot numbers and locations and best subset regressions. The computational features of the package were discussed, and several applications were given.    

\section{Acknowledgments}
Mo Li thanks the Louisiana Board of Regents Support Fund (BoRSF) Research Competitiveness Subprogram LEQSF(2025-28)-RD-A-18 for partial support;   Xueheng Shi thanks the University of Nebraska-Lincoln Grant ARD-2162251011 for partial support.


%



\vspace{1cm}

\newpage

\bibliographystyle{plainnat}
\bibliography{arxiv.bib}

\end{document}